\documentclass[prl,twocolumn,amsmath,amssymb,showpacs]{revtex4}
\usepackage{bm}
\usepackage{graphicx}

\begin{document}


\newcommand{\dtwomin}{D^2_\text{min}}
\newcommand{\eg}{\emph{e.g.}\ }
\newcommand{\ie}{\emph{i.e.}\ }

\title{Mechanical and microscopic properties of the reversible plastic regime in a 2D jammed material}

\author{Nathan~C.~Keim}
\email{nkeim@seas.upenn.edu}
\affiliation{Department of Mechanical Engineering and Applied Mechanics, University of Pennsylvania, Philadelphia, PA 19104}
\author{Paulo~E.~Arratia}
\email{parratia@seas.upenn.edu}
\affiliation{Department of Mechanical Engineering and Applied Mechanics, University of Pennsylvania, Philadelphia, PA 19104}


\date{\today}

\begin{abstract}

At the microscopic level, plastic flow of a jammed, disordered material consists of a series of particle rearrangements that cannot be reversed by subsequent deformation. An infinitesimal deformation of the same material has no rearrangements. Yet between these limits, there may be a self-organized plastic regime with rearrangements, but with no net change upon reversing a deformation. We measure the oscillatory response of a jammed interfacial material, and directly observe rearrangements that couple to bulk stress and dissipate energy, but do not always give rise to global irreversibility.

\end{abstract}

\maketitle

The mechanical properties of disordered (amorphous) materials far from equilibrium --- from sand, to plastics, to ice cream --- continue to elude comprehensive understanding~\cite{Larson:1998ud,Chen:2008kg,Chen:2010jn}. These materials typically feature many particles (\eg droplets, atoms, or grains) that are crowded together in close contact, and are both \emph{jammed} so that each particle is fully constrained by its neighbors, and \emph{disordered} so that these constraints vary greatly among particles, and crystalline order rarely extends beyond several particle diameters~\cite{VanHecke:2010go}. A sufficiently large imposed stress may cause these materials to flow plastically as would a viscous liquid, permanently changing the equilibrium arrangement --- the microstructure --- of the particles. Plastic flow, and the process of yielding that initiates it, are governed by local structural relaxations in which one particle squeezes past another, relieving nearby stresses and dissipating energy. These relaxations and many other behaviors are common to materials on a wide range of length scales and with varying microscopic physics, but the way specific microscopic processes organize and give rise to macroscopic behaviors --- the material's bulk rheology --- is still not well-understood~\cite{Gopal:1995bg, Schall:2007fd, Manning:2011ha,Falk:2011co}.

\begin{figure}
\includegraphics[width=2.5in]{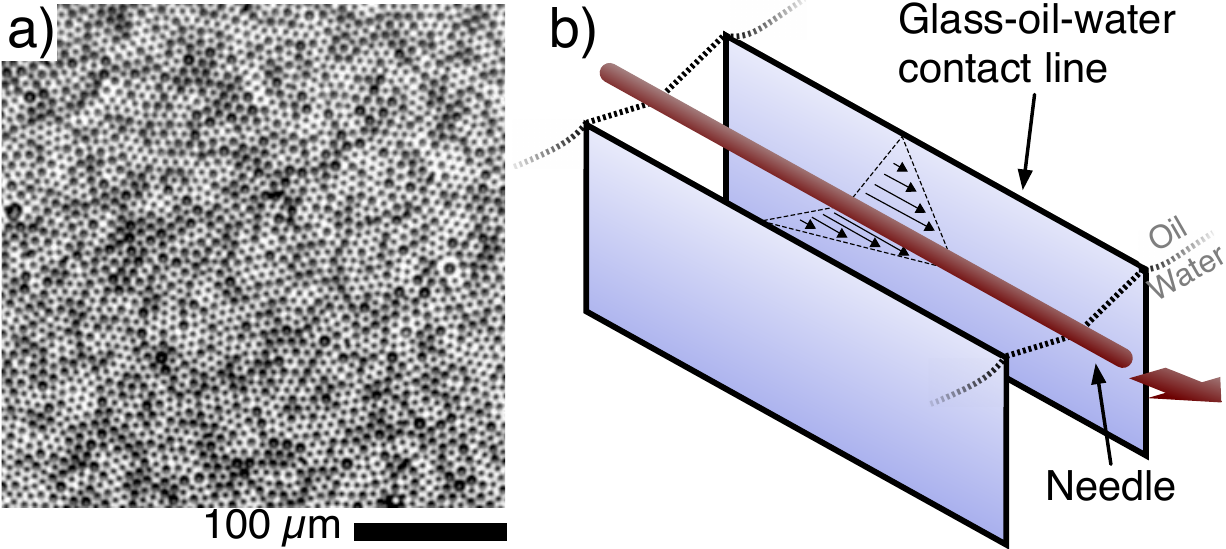}
\centering
\caption{(color online) Material and apparatus.
\textbf{(a)} View from above of bulk material: mutually repulsive polystyrene microspheres adsorbed at oil-water interface.
\textbf{(b)} Interfacial stress rheometer apparatus. The interfacial material is pinned on glass walls; a needle is embedded in the material between them, and is magnetically forced. Velocity profile is sketched.
\label{fig:apparatus}}
\end{figure}

If the timescale of structural relaxation is much shorter than any global timescale of deformation (\eg a period of driving or the inverse strain rate $\dot \gamma^{-1}$), we can describe changes to microstructure in terms of discrete, local plastic rearrangements, which are a key feature of the shear transformation zone (STZ) picture of plasticity~\cite{Falk:1998wm,Falk:2011co}. Under steady shear, the piling-on of these events, each of which traverses a barrier between two local minima in potential energy, ensures that the initial microstructure can never be recovered out of a vast landscape of metastable states. However, it is believed that individually and in isolation, many if not all plastic rearrangements can be reverted by applying a reverse stress~\cite{Argon:1979iy,lundberg08,Falk:1998wm,Falk:2011co}. Furthermore, when a material is deformed cyclically with sufficiently small amplitude, recent simulations and experiments have observed that reversing the deformation may reverse virtually all changes, returning the entire material to its original state~\cite{Slotterback:2012fa,Ren:2013cp,Priezjev:2013hp,Keim:2013je,Regev:2013preprint,Schreck:2013preprint,Fiocco:2013ds}. Viewed stroboscopically (once per cycle), the microstructure is static. This poses new questions for a complete description of material response: When rearrangements are stroboscopically invisible, do they meaningfully affect bulk rheology? What are their characteristics? Could they clarify the yielding transition, when bulk properties change and the material becomes stroboscopically dynamic?~\cite{Keim:2013je,Regev:2013preprint,Fiocco:2013ds,Petekidis:2002by}

Here, we examine in detail the rearrangements in a cyclically-sheared jammed material, in experiments in which it self-organizes to a steady state that is stroboscopically static~\cite{Keim:2013je}. The material is a monolayer of repulsive microspheres adsorbed at an oil-water interface, for which we simultaneously measure mechanical response (rheology) and image many ($5.6 \times 10^4$) individual particles. We find that even when the deformation is globally reversible, local rearrangements are plastic, displaying hysteresis and altering rheology. The former is a sign that the self-organized steady state is in fact a limit cycle, as found in many other nonlinear dynamical systems~\cite{Strogatz:1994tz, Regev:2013preprint, Schreck:2013preprint}. This reversible plasticity vanishes at small strain amplitude, and is gradually overwhelmed by irreversibility as the yielding transition is surpassed. Our findings strongly suggest that microscopic rearrangements and bulk plasticity are necessary but not sufficient for irreversibility.

Our model material is a mixture (equal parts by number) of 4.1 and 5.6 $\mu$m-diameter sulfate latex (polystyrene) particles (Invitrogen; nominal diameters 4 and 6 $\mu$m) adsorbed at a water-decane interface with area fraction $\phi \sim 0.43$. The particles do not touch, but their electrostatic dipole-dipole repulsion~\cite{Masschaele:2010da} results in a stable, soft (\ie readily deformable) jammed material (Fig.~\ref{fig:apparatus}a); the particles' large sizes and strong repulsion make thermal motion negligible. This material is subjected to a linear shear deformation in an interfacial stress rheometer (ISR)~\cite{suppmat,Brooks:1999ky,Reynaert:2008dm}. As shown in Fig.~\ref{fig:apparatus}b, a magnetized needle is placed on the material to be studied, in an open channel formed by 2 vertical glass walls. An electromagnet forces the needle, creating a uniform shear stress $\sigma(t)$ on the material between the needle and the walls. We measure material rheology by observing the needle's motion [expressed as strain $\gamma(t)$] under oscillatory stress. 

Deformation in experiments is quasistatic and rearrangements are discrete, insofar as the timescale for a rearrangement to complete ($\sim 0.5$~s) is much shorter than the shortest driving period (5~s) or largest inverse strain rate ($\dot \gamma^{-1} = 20$~s). We also require that the boundary conditions in the 3rd dimension be approximately stress-free --- that typical forces in the plane of the material are much stronger than viscous drag from the liquid bath~\cite{Reynaert:2008dm}. This ratio is the Boussinesq number $Bq = |\eta^*| a / \eta_l$, where $\eta^*$ is the material's observed complex viscosity, $a = 230$~$\mu$m is the needle diameter, and $\eta_l \simeq 10^{-3}$~Pa~s is the oil and water viscosity. Here $Bq \sim 10^2$ and so our experiments are nearly 2D. Further details of the material and apparatus are found in the Supplemental Materials~\cite{suppmat}. For each experiment, we prepare the material with 6 cycles of shearing at large amplitude ($\gamma_0 \sim 0.5$), then stop. Resuming at smaller $\gamma_0$ starts a transient relaxation to a steady state.

At each cycle of driving during the experiment, we can measure \emph{total} (peak-to-peak) change in microstructure by comparing particle positions at a minimum of global strain $\gamma(t_\text{min})$ with those at the following maximum $\gamma(t_\text{max})$. \emph{Irreversible} change is measured stroboscopically, by sampling at times $(t_\text{max} + t_\text{min} \pm 2\pi \omega^{-1}) / 2$, so that we compare the beginning and end of a full period of driving that straddles $t_\text{min}$ and $t_\text{max}$. Wherever there is no irreversible change to microstructure, any total change in that same cycle is by definition reversible.

\begin{figure*}
\includegraphics[width=17.1cm]{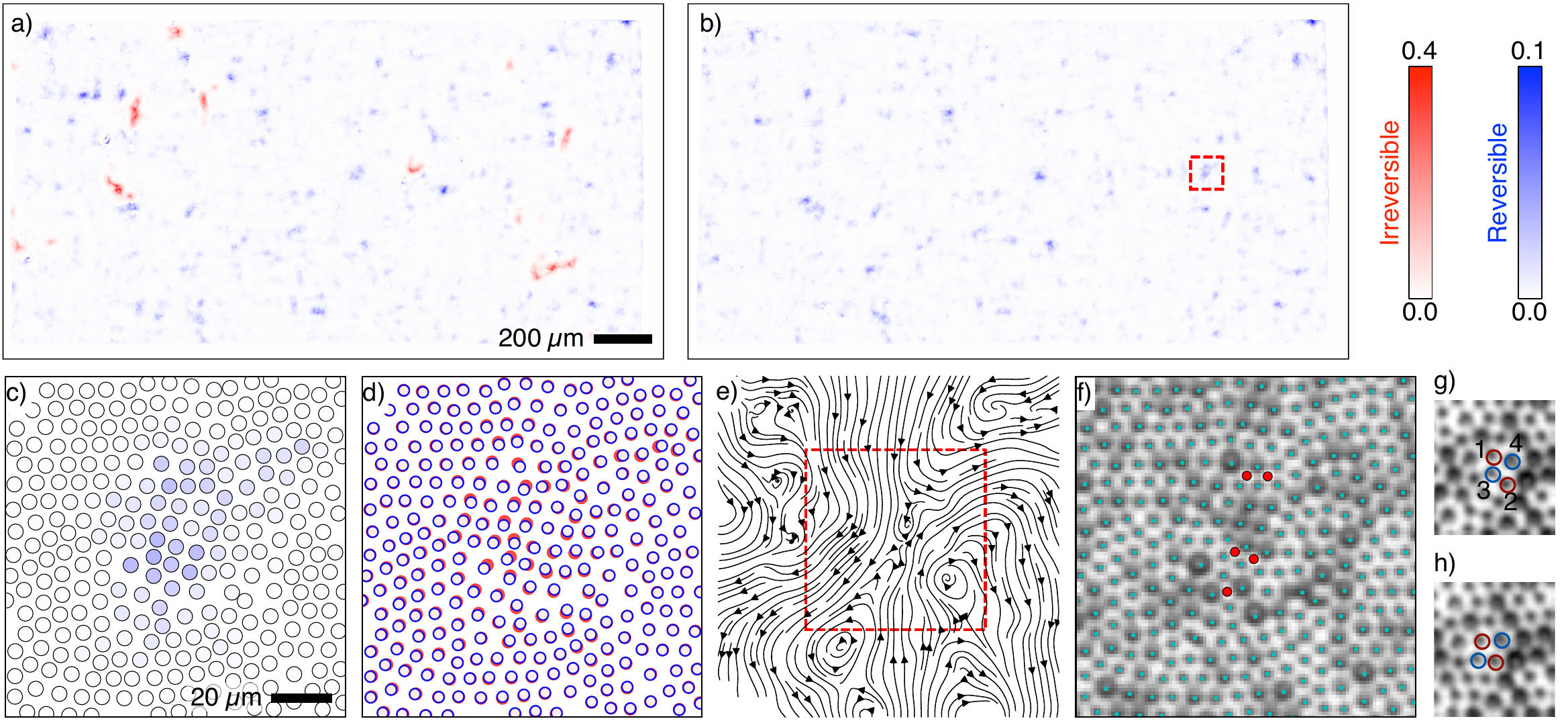}
\centering
\caption{(color online) Local deformation in a plastic event. 
\textbf{(a,b)} Total (\ie peak-to-peak, blue) and irreversible (\ie stroboscopic, overlaid in red) $\dtwomin$ for cycles (a) 8 and (b) 20 of shear at $\gamma_0 = 0.020$, showing clusters of non-affine deformation. One reversible cluster in (b) is boxed, and shown in (c--f). The magnetic needle is at the top of the image; the fixed wall is at the bottom.
\textbf{(c)} Detail of a reversible cluster, showing the $\dtwomin$ of individual particles. Color scale is the same as in (a,b).
\textbf{(d)} Local relative displacement of particles in (c) at the minimum (red) and subsequent maximum (blue open circles) of $\gamma$, subtracting motion of neighbors within $10a$.
\textbf{(e)} Streamlines computed from displacements (subtracting motion of neighbors within $40a$); square outline is region of (c,d,f). The hyperbolic character of the displacements is evident in the far-field.
\textbf{(f)} Micrograph with particle centers (small dots) and the centroid of the 4 particles in each T1 rearrangement (large dots) marked.
\textbf{(g,h)} Sequence illustrating a T1 rearrangement. Particles 3 and 4 begin as nearest neighbors but are separated in (h).
\label{fig:activitymap}
}
\end{figure*}

\begin{figure}
\includegraphics[width=2.4in]{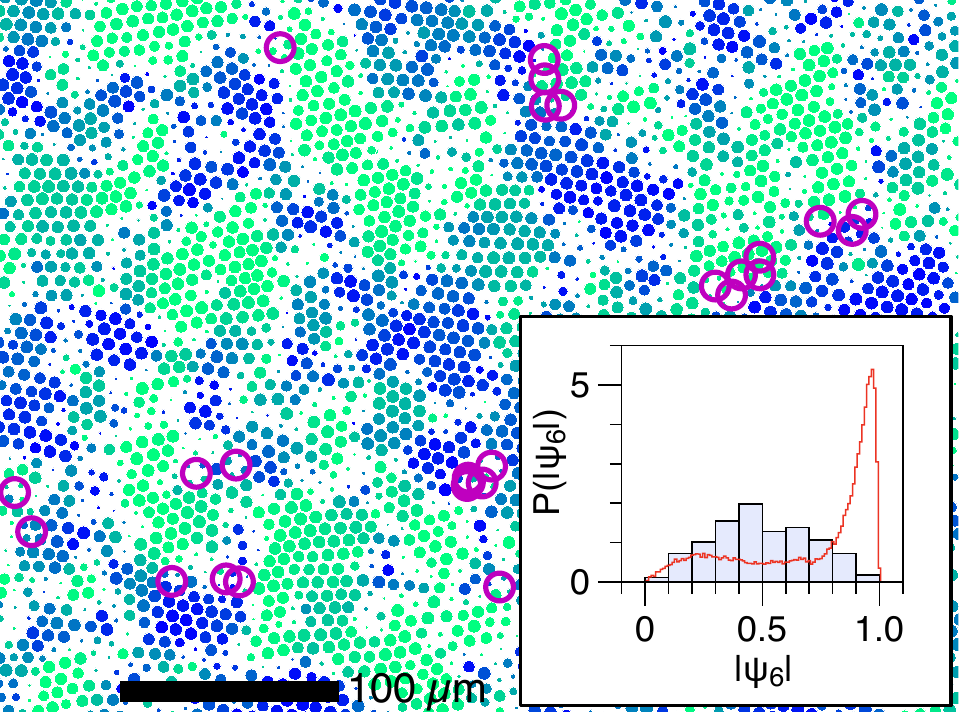}
\centering
\caption{(color online) Rearrangements in portion of material at $\gamma_0 = 0.02$ during transient (cycle 15; chosen to obtain more events than in steady state). Each particle is shown as solid dot with size representing the extent of local crystalline ordering $|\psi_6|$; minimum and maximum size signify $|\psi_6| \simeq 0.1$ and 1. Color is solely to show differences in lattice director. The centroids of total T1 events (see Fig.~\ref{fig:activitymap}f) are shown as large open circles.
\textbf{Inset:} Histograms of $|\psi_6|$ in reversible steady state (cycle 20). Curve: all particles. Shaded bars: particles involved in T1 events (555 out of $5.6 \times 10^4$). Dot positions and $\psi_6$ are for $\gamma \simeq \langle \gamma \rangle$.
\label{fig:grainboundaries}}
\end{figure}

Figure~\ref{fig:activitymap} shows changes to microstructure in single cycles of deformation, for the entire system and for a single region. Panels (a,b,c) detect rearrangements with the quantity $\dtwomin$, computed between 2 instants by measuring how much each particle and its 2 nearest ``shells'' of neighbors move \emph{unlike} a continuous elastic solid; it is the mean squared residual displacement after subtracting the best affine transformation~\cite{Falk:1998wm}. $\dtwomin$ is normalized by the square of the typical interparticle spacing, $a \simeq 6.8$~$\mu$m; details are given in the Supplemental Materials~\cite{suppmat}. Figures~\ref{fig:activitymap}(a,b) illustrate evolution to a reversible steady state in which rearrangements occur, but are always reversed by the end of each cycle; movies SM1--3 show the full evolution at 3 strain amplitudes~\cite{suppmat}. We set a threshold $D^2_0 = 0.015$, corresponding to a disturbance $\sim 0.1a \simeq 1$~pixel, and comparable with a value used for simulations of disordered solids~\cite{Falk:1998wm}. Most particles in Fig.~\ref{fig:activitymap}b have $\dtwomin \lesssim 10^{-3}$, while those with $\dtwomin \ge D^2_0$ are in clusters of $\lesssim 20$ particles, with median size $\sim 5$ particles.

We may also measure change to microstructure as the displacement of a particle relative to the material around it (Fig.~\ref{fig:activitymap}d). The resulting computed streamlines (Fig.~\ref{fig:activitymap}e) resemble the flow at a hyperbolic point in an incompressible fluid, consistent with the geometry of a single plastic event measured in sheared dry foams by Kabla and Debregas~\cite{Kabla:2003it}, and modeled by Picard \textit{et al.}~\cite{Picard:2004cx} for an otherwise elastic incompressible medium. Finally, rearranging particles lose and gain nearest neighbors, a process discretized as T1 events~\cite{Weaire:1984gu} in Fig.~\ref{fig:activitymap}f, and Movie SM4~\cite{suppmat}. Details of these computations are in the Supplemental Materials~\cite{suppmat}.

We find that the locations of rearrangements are not predicted by static material structure, such as local number density, presence of anomalously large or small particles, or number of neighbors. However, we do see a difference between more- and less-ordered regions. The bond order parameter magnitude $|\psi_6|$ measures the degree to which each particle's neighbors are spaced $60^\circ$ apart (details in Supplemental Materials \cite{suppmat}); Fig.~\ref{fig:grainboundaries} shows that the material has of regions of crystalline order with scale $\sim 5a$, and thick interstitial ``grain boundaries.'' Particles involved in plasticity are disproportionately in the latter, strongly suggesting that the material's response is dominated by disorder.

\begin{figure}
\includegraphics[width=2.4in]{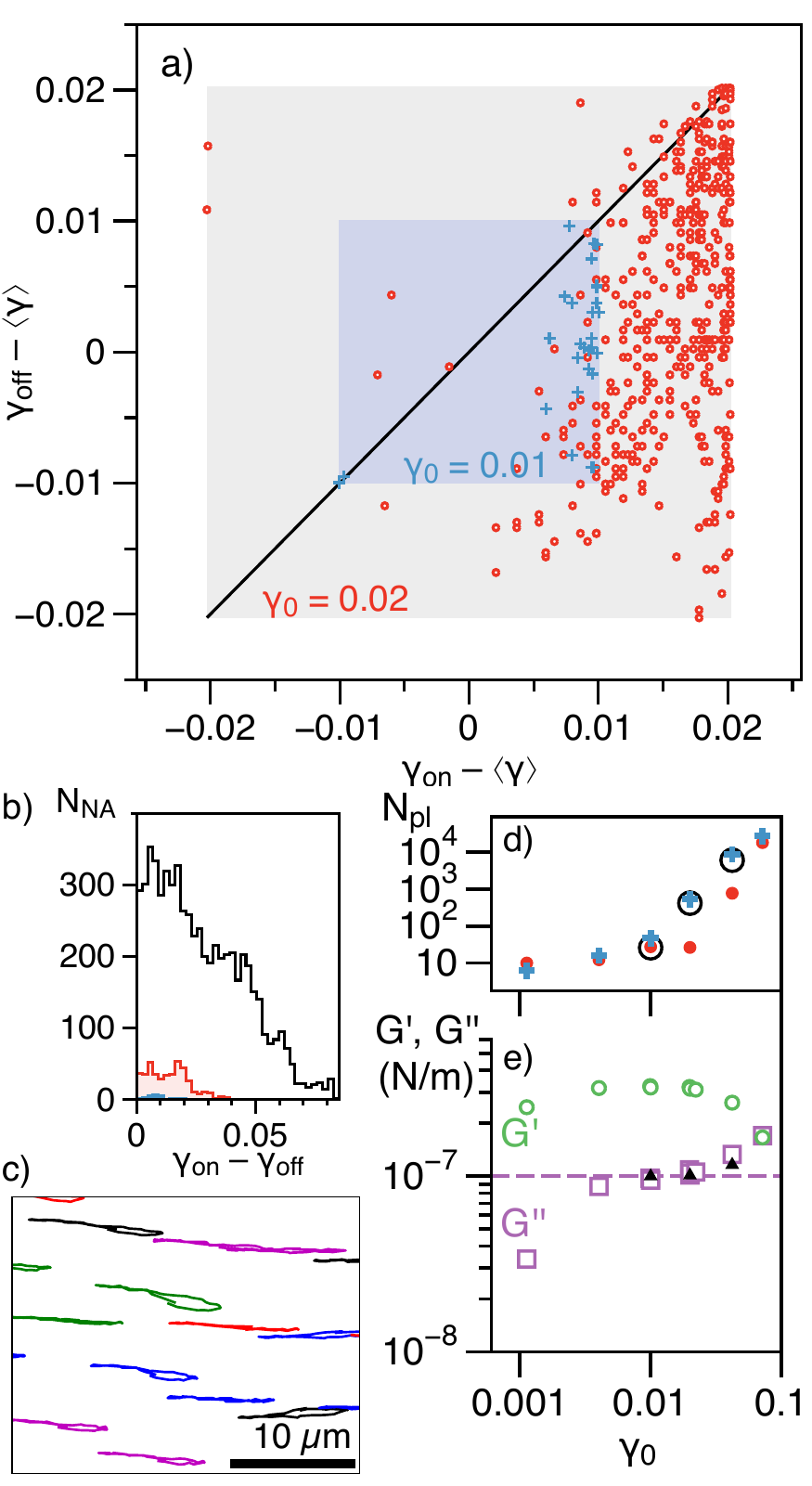}
\centering
\caption{(color online) Hysteresis of rearrangements.
\textbf{(a)} Global strain $\gamma$ at which individual particles rearrange ($\dtwomin \ge 0.015$) and reverse, in steady state for $\gamma_0 = 0.01$ ($+$) and $\gamma_0 = 0.02$ ($\circ$). Shaded regions show limits of $\gamma$. The farther an event falls from the diagonal line, the more hysteretic it is. 
\textbf{(b)} Histogram of hysteresis at $\gamma_0 = 0.04$ (open, black curve), $0.02$ (shaded, red), $0.01$ (solid, blue). 
\textbf{(c)} Particle trajectories break time-reversal symmetry. A rearranging portion of the system is shown for one cycle in the steady state ($\gamma_0 = 0.02$). Colors distinguish the particles. The effect of microscope vibration is reduced by subtracting average $y$ (here, vertical) motion of entire visible system.
\textbf{(d)} Number of particles $N_\text{pl}$ in plastic rearrangements in steady state, as function of $\gamma_0$. Points show number (out of $5.6 \times 10^4$) with total ($+$) or irreversible ($\bullet$) $\dtwomin \ge 0.015$. Hysteretic particles ($\circ$) have $\gamma_\text{on} - \gamma_\text{off}$ exceeding the largest change in $\gamma$ between video frames.
\textbf{(e)} Oscillatory rheology. $\blacktriangle$: Estimated enhancement of $G''$ above zero-plasticity level (dashed line), based on microstructure (see text).
\label{fig:hysteresis}}
\end{figure}

We now verify that these rearrangements are a form of plasticity, dissipating energy and coupling to bulk stress; this does not necessarily follow from non-affine deformation alone~\cite{VanHecke:2010go,Zaccone:2011cm}. A plastic rearrangement is caused by a local buildup of (elastic) stress as the whole material is sheared; an opposing buildup is required to reverse it. Such events appear hysteretic, turning ``on'' during forward shear at a global strain $\gamma_\text{on}$, and ``off'' during reverse shear at $\gamma_\text{off}$, with $\gamma_\text{on} - \gamma_\text{off} > 0$ as a proxy for the activating stress. 

Figure~\ref{fig:hysteresis}a shows hysteresis in a single cycle, using $\gamma(t) = \gamma_\text{min}$ as the undeformed state. Using the threshold $D^2_0 = 0.015$, we obtain a $\gamma_\text{on}$ at the last video frame for which a particle's $\dtwomin < D^2_0$, and $\gamma_\text{off}$ at the last frame with $\dtwomin \ge D^2_0$. We require ``on'' and ``off'' to be in the first and second halves of the cycle respectively, and $\dtwomin \ge D^2_0$ for at least 50\% of the intervening frames (for most events this approaches 100\%). At the extreme, some rearrangements activate at $\sim \gamma_\text{max}$ but reverse at $\sim \gamma_\text{min}$. Figure~\ref{fig:hysteresis}b shows that hysteretic plasticity grows dramatically in abundance and strength as $\gamma_0$ is increased. Hysteresis breaks time-reversal symmetry, as also seen in the looped trajectories of Fig.~\ref{fig:hysteresis}c, and it locally makes strain a multiple-valued function of stress. These behaviors are inconsistent with purely elastic deformation and consistent with plasticity as described by STZ theory~\cite{Falk:1998wm,lundberg08,Falk:2011co}.

We can now connect our simultaneous observations of rheology and microscopic behavior in the steady state. Numbers of rearranging particles ($\dtwomin \ge D^2_0$) averaged over the final 3 cycles of each movie at various $\gamma_0$ are plotted in Fig.~\ref{fig:hysteresis}d; behavior changes little over at least 4 cycles. To measure rheology in Fig.~\ref{fig:hysteresis}e, we model stress as the real part of $(G' + iG'') \gamma$, with $\gamma = e^{i \omega t}$, where $\omega$ is the angular frequency of driving; this gives a storage modulus $G'$, measuring elastic character, and loss modulus $G''$, measuring viscous or plastic character. As discussed above, $\gamma_\text{on} - \gamma_\text{off}$ is a proxy for the local stress $\sigma_\text{pl}$ causing the rearrangement, and for its contribution to dissipation (\ie to $G''$). Using the relation for dissipation per unit area per cycle, $w_\text{cyc} = \pi \gamma_0^2 G''$, and the data in Fig.~\ref{fig:hysteresis}e, we can estimate the plastic contribution to $G''$ in a reversible or mostly-reversible steady state ($\gamma_0 \le 0.04$),
\begin{equation}
G_\text{pl}'' = \frac{2}{\pi \gamma_0^2 A} \sum_i  G' a^2 (\gamma_\text{on}^i - \gamma_\text{off}^i)^2
\end{equation}
where $A$ is the area of observations, 2 refers to each particle switching twice per cycle, and the sum estimates the elastic energy built up and then dissipated, for each particle in Fig.~\ref{fig:hysteresis}b. This estimate, made by choosing $D^2_0$ only, is shown in Fig.~\ref{fig:hysteresis}e. It is of the same order as the actual increase in $G''$ at $\gamma_0 = 0.04$. 

Using simultaneous bulk rheometry and particle tracking under shear, we have studied the nature and mechanical role of microscopic plastic events in a soft jammed material. This material can evolve to a steady state in which mechanical response is primarily elastic and microstructure is unchanged by each cycle~\cite{Keim:2013je}, and yet some particles rearrange plastically during deformation. This regime is due to a stable population of rearrangements, comprising just $\sim$1\% of particles, suggesting that to reliably observe it, $\gtrsim$$10^3$ particles must be studied. It is reminiscent of a limit cycle, a closed trajectory in phase space that a nonlinear system may evolve toward~\cite{Strogatz:1994tz}, and which describes simulations of cyclically-sheared athermal frictionless jammed~\cite{Regev:2013preprint} and unjammed particles~\cite{Schreck:2013preprint}. Limit cycles break time-reversal symmetry, as seen in the looped trajectories of Fig.~\ref{fig:hysteresis}c, and so are much more general than the linear dynamics of the reversible steady state in dilute non-Brownian suspensions~\cite{Corte:2008tp}. Our finding of limit cycles may depend weakly on the duration of the experiment, in that thermal or mechanical noise could cause sporadic further relaxations~\cite{Richard:2005dj,Nguyen:2011hm}.

Considering the results discussed in this work, both in our experiments and published elsewhere~\cite{lundberg08,Petekidis:2002by,Priezjev:2013hp,Keim:2013je,Regev:2013preprint,Schreck:2013preprint,Fiocco:2013ds}, we see 3 regimes of steady-state cyclic deformation: 
(1) Far below yielding ($\gamma_0 \ll \gamma_y$), response is truly elastic and time-reversible, with no rearrangements. Nonetheless, some particle motions may be non-affine due to disorder~\cite{VanHecke:2010go,Zaccone:2011cm}.
(2) As $\gamma_0 \to \gamma_y$, microscopic plasticity grows rapidly. Rheological response is still dominated by elasticity ($G' \gg G''$), and the material is stroboscopically static~\cite{Priezjev:2013hp,Keim:2013je,Regev:2013preprint,Schreck:2013preprint,Fiocco:2013ds}, but time-reversibility is broken~\cite{Regev:2013preprint,Schreck:2013preprint}. Plasticity contributes to $G''$ but may not dominate. 
(3) $\gamma_0 = \gamma_y^\text{micro}$ marks the appearance of irreversible plasticity in the steady state and is a clearly-defined yielding transition~\cite{Keim:2013je,Regev:2013preprint,Fiocco:2013ds,Petekidis:2002by}. Much of the system may be nonetheless reversible in a given cycle (see Fig.~\ref{fig:hysteresis}d or Movie SM3)~\cite{Hebraud:1997ef,lundberg08,suppmat}. On the other hand, the rheological yielding transition, wherein $G''$ increases and elasticity declines, is gradual; at the microscopic level it is due to both reversible and irreversible plasticity. 

Our work shows that in an experimental jammed material, plasticity and irreversibility can become decoupled in the steady-state oscillatory response: the material can host many microscopic plastic rearrangements that couple to the bulk stress and dissipate energy, yet do not give rise to global irreversibility. This strongly suggests a qualitative difference between microstructural yielding (the transition to irreversibility) and rheological yielding: rearrangements and bulk plasticity are necessary but not sufficient for irreversibility. Differences between the restricted, self-organized STZ-like rearrangements of the reversible steady state, and a more general population under steady shear, may shed light on models of STZ populations~\cite{Falk:2011co}, or other measures of static and dynamical structure~\cite{Chen:2008kg,Manning:2011ha,Chen:2011fa}.

We thank John Brady, Andrea Liu, Martin van Hecke, and Ye Xu for helpful discussions. This work was supported by the Penn NSF MRSEC (DMR-1120901).

\bibliography{references,references-suppmat}

\begin{thebibliography}{33}
\expandafter\ifx\csname natexlab\endcsname\relax\def\natexlab#1{#1}\fi
\expandafter\ifx\csname bibnamefont\endcsname\relax
  \def\bibnamefont#1{#1}\fi
\expandafter\ifx\csname bibfnamefont\endcsname\relax
  \def\bibfnamefont#1{#1}\fi
\expandafter\ifx\csname citenamefont\endcsname\relax
  \def\citenamefont#1{#1}\fi
\expandafter\ifx\csname url\endcsname\relax
  \def\url#1{\texttt{#1}}\fi
\expandafter\ifx\csname urlprefix\endcsname\relax\def\urlprefix{URL }\fi
\providecommand{\bibinfo}[2]{#2}
\providecommand{\eprint}[2][]{\url{#2}}

\bibitem[{\citenamefont{Larson}(1998)}]{Larson:1998ud}
\bibinfo{author}{\bibfnamefont{R.~G.} \bibnamefont{Larson}},
  \emph{\bibinfo{title}{{The Structure and Rheology of Complex Fluids}}}
  (\bibinfo{publisher}{Oxford}, \bibinfo{year}{1998}).

\bibitem[{\citenamefont{Chen}(2008)}]{Chen:2008kg}
\bibinfo{author}{\bibfnamefont{M.}~\bibnamefont{Chen}}, \bibinfo{journal}{Annu.
  Rev. Mater. Res.} \textbf{\bibinfo{volume}{38}}, \bibinfo{pages}{445}
  (\bibinfo{year}{2008}).

\bibitem[{\citenamefont{Chen et~al.}(2010)\citenamefont{Chen, Wen, Janmey,
  Crocker, and Yodh}}]{Chen:2010jn}
\bibinfo{author}{\bibfnamefont{D.~T.~N.} \bibnamefont{Chen}},
  \bibinfo{author}{\bibfnamefont{Q.}~\bibnamefont{Wen}},
  \bibinfo{author}{\bibfnamefont{P.~A.} \bibnamefont{Janmey}},
  \bibinfo{author}{\bibfnamefont{J.~C.} \bibnamefont{Crocker}},
  \bibnamefont{and} \bibinfo{author}{\bibfnamefont{A.~G.} \bibnamefont{Yodh}},
  \bibinfo{journal}{Annu. Rev. Condens. Matter Phys.}
  \textbf{\bibinfo{volume}{1}}, \bibinfo{pages}{301} (\bibinfo{year}{2010}).

\bibitem[{\citenamefont{van Hecke}(2010)}]{VanHecke:2010go}
\bibinfo{author}{\bibfnamefont{M.}~\bibnamefont{van Hecke}},
  \bibinfo{journal}{J. Phys: Cond. Matter} \textbf{\bibinfo{volume}{22}},
  \bibinfo{pages}{3101} (\bibinfo{year}{2010}).

\bibitem[{\citenamefont{Gopal and Durian}(1995)}]{Gopal:1995bg}
\bibinfo{author}{\bibfnamefont{A.~D.} \bibnamefont{Gopal}} \bibnamefont{and}
  \bibinfo{author}{\bibfnamefont{D.~J.} \bibnamefont{Durian}},
  \bibinfo{journal}{Phys. Rev. Lett.} \textbf{\bibinfo{volume}{75}},
  \bibinfo{pages}{2610} (\bibinfo{year}{1995}).

\bibitem[{\citenamefont{Schall et~al.}(2007)\citenamefont{Schall, Weitz, and
  Spaepen}}]{Schall:2007fd}
\bibinfo{author}{\bibfnamefont{P.}~\bibnamefont{Schall}},
  \bibinfo{author}{\bibfnamefont{D.~A.} \bibnamefont{Weitz}}, \bibnamefont{and}
  \bibinfo{author}{\bibfnamefont{F.}~\bibnamefont{Spaepen}},
  \bibinfo{journal}{Science} \textbf{\bibinfo{volume}{318}},
  \bibinfo{pages}{1895} (\bibinfo{year}{2007}).

\bibitem[{\citenamefont{Manning and Liu}(2011)}]{Manning:2011ha}
\bibinfo{author}{\bibfnamefont{M.~L.} \bibnamefont{Manning}} \bibnamefont{and}
  \bibinfo{author}{\bibfnamefont{A.~J.} \bibnamefont{Liu}},
  \bibinfo{journal}{Phys. Rev. Lett.} \textbf{\bibinfo{volume}{107}},
  \bibinfo{pages}{108302} (\bibinfo{year}{2011}).

\bibitem[{\citenamefont{Falk and Langer}(2011)}]{Falk:2011co}
\bibinfo{author}{\bibfnamefont{M.~L.} \bibnamefont{Falk}} \bibnamefont{and}
  \bibinfo{author}{\bibfnamefont{J.~S.} \bibnamefont{Langer}},
  \bibinfo{journal}{Annu. Rev. Condens. Matter Phys.}
  \textbf{\bibinfo{volume}{2}}, \bibinfo{pages}{353} (\bibinfo{year}{2011}).

\bibitem[{\citenamefont{Falk and Langer}(1998)}]{Falk:1998wm}
\bibinfo{author}{\bibfnamefont{M.~L.} \bibnamefont{Falk}} \bibnamefont{and}
  \bibinfo{author}{\bibfnamefont{J.~S.} \bibnamefont{Langer}},
  \bibinfo{journal}{Phys. Rev. E} \textbf{\bibinfo{volume}{57}},
  \bibinfo{pages}{7192} (\bibinfo{year}{1998}).

\bibitem[{\citenamefont{Argon}(1979)}]{Argon:1979iy}
\bibinfo{author}{\bibfnamefont{A.~S.} \bibnamefont{Argon}},
  \bibinfo{journal}{Acta Metallurgica} \textbf{\bibinfo{volume}{27}},
  \bibinfo{pages}{47} (\bibinfo{year}{1979}).

\bibitem[{\citenamefont{Lundberg et~al.}(2008)\citenamefont{Lundberg, Krishan,
  Xu, O'Hern, and Dennin}}]{lundberg08}
\bibinfo{author}{\bibfnamefont{M.}~\bibnamefont{Lundberg}},
  \bibinfo{author}{\bibfnamefont{K.}~\bibnamefont{Krishan}},
  \bibinfo{author}{\bibfnamefont{N.}~\bibnamefont{Xu}},
  \bibinfo{author}{\bibfnamefont{C.~S.} \bibnamefont{O'Hern}},
  \bibnamefont{and} \bibinfo{author}{\bibfnamefont{M.}~\bibnamefont{Dennin}},
  \bibinfo{journal}{Phys. Rev. E} \textbf{\bibinfo{volume}{77}},
  \bibinfo{pages}{041505} (\bibinfo{year}{2008}).

\bibitem[{\citenamefont{Slotterback et~al.}(2012)\citenamefont{Slotterback,
  Mailman, Ronaszegi, van Hecke, Girvan, and Losert}}]{Slotterback:2012fa}
\bibinfo{author}{\bibfnamefont{S.}~\bibnamefont{Slotterback}},
  \bibinfo{author}{\bibfnamefont{M.}~\bibnamefont{Mailman}},
  \bibinfo{author}{\bibfnamefont{K.}~\bibnamefont{Ronaszegi}},
  \bibinfo{author}{\bibfnamefont{M.}~\bibnamefont{van Hecke}},
  \bibinfo{author}{\bibfnamefont{M.}~\bibnamefont{Girvan}}, \bibnamefont{and}
  \bibinfo{author}{\bibfnamefont{W.}~\bibnamefont{Losert}},
  \bibinfo{journal}{Phys. Rev. E} \textbf{\bibinfo{volume}{85}},
  \bibinfo{pages}{021309} (\bibinfo{year}{2012}).

\bibitem[{\citenamefont{Ren et~al.}(2013)\citenamefont{Ren, Dijksman, and
  Behringer}}]{Ren:2013cp}
\bibinfo{author}{\bibfnamefont{J.}~\bibnamefont{Ren}},
  \bibinfo{author}{\bibfnamefont{J.~A.} \bibnamefont{Dijksman}},
  \bibnamefont{and} \bibinfo{author}{\bibfnamefont{R.~P.}
  \bibnamefont{Behringer}}, \bibinfo{journal}{Phys. Rev. Lett.}
  \textbf{\bibinfo{volume}{110}}, \bibinfo{pages}{018302}
  (\bibinfo{year}{2013}).

\bibitem[{\citenamefont{Priezjev}(2013)}]{Priezjev:2013hp}
\bibinfo{author}{\bibfnamefont{N.~V.} \bibnamefont{Priezjev}},
  \bibinfo{journal}{Phys. Rev. E} \textbf{\bibinfo{volume}{87}},
  \bibinfo{pages}{052302} (\bibinfo{year}{2013}).

\bibitem[{\citenamefont{Keim and Arratia}(2013)}]{Keim:2013je}
\bibinfo{author}{\bibfnamefont{N.~C.} \bibnamefont{Keim}} \bibnamefont{and}
  \bibinfo{author}{\bibfnamefont{P.~E.} \bibnamefont{Arratia}},
  \bibinfo{journal}{Soft Matter} \textbf{\bibinfo{volume}{9}},
  \bibinfo{pages}{6222} (\bibinfo{year}{2013}).

\bibitem[{\citenamefont{Regev et~al.}(2013)\citenamefont{Regev, Lookman, and
  Reichhardt}}]{Regev:2013preprint}
\bibinfo{author}{\bibfnamefont{I.}~\bibnamefont{Regev}},
  \bibinfo{author}{\bibfnamefont{T.}~\bibnamefont{Lookman}}, \bibnamefont{and}
  \bibinfo{author}{\bibfnamefont{C.}~\bibnamefont{Reichhardt}}
  (\bibinfo{year}{2013}), \bibinfo{note}{{arXiv:1301.7479}}.

\bibitem[{\citenamefont{Schreck et~al.}(2013)\citenamefont{Schreck, Hoy,
  Shattuck, and O'Hern}}]{Schreck:2013preprint}
\bibinfo{author}{\bibfnamefont{C.~F.} \bibnamefont{Schreck}},
  \bibinfo{author}{\bibfnamefont{R.~S.} \bibnamefont{Hoy}},
  \bibinfo{author}{\bibfnamefont{M.~D.} \bibnamefont{Shattuck}},
  \bibnamefont{and} \bibinfo{author}{\bibfnamefont{C.~S.} \bibnamefont{O'Hern}}
  (\bibinfo{year}{2013}), \bibinfo{note}{{arXiv:1301.7492v1}}.

\bibitem[{\citenamefont{Fiocco et~al.}(2013)\citenamefont{Fiocco, Foffi, and
  Sastry}}]{Fiocco:2013ds}
\bibinfo{author}{\bibfnamefont{D.}~\bibnamefont{Fiocco}},
  \bibinfo{author}{\bibfnamefont{G.}~\bibnamefont{Foffi}}, \bibnamefont{and}
  \bibinfo{author}{\bibfnamefont{S.}~\bibnamefont{Sastry}},
  \bibinfo{journal}{Phys. Rev. E} \textbf{\bibinfo{volume}{88}},
  \bibinfo{pages}{020301} (\bibinfo{year}{2013}).

\bibitem[{\citenamefont{Petekidis et~al.}(2002)\citenamefont{Petekidis,
  Moussa{\"\i}d, and Pusey}}]{Petekidis:2002by}
\bibinfo{author}{\bibfnamefont{G.}~\bibnamefont{Petekidis}},
  \bibinfo{author}{\bibfnamefont{A.}~\bibnamefont{Moussa{\"\i}d}},
  \bibnamefont{and} \bibinfo{author}{\bibfnamefont{P.~N.} \bibnamefont{Pusey}},
  \bibinfo{journal}{Phys. Rev. E} \textbf{\bibinfo{volume}{66}},
  \bibinfo{pages}{051402} (\bibinfo{year}{2002}).

\bibitem[{\citenamefont{Strogatz}(1994)}]{Strogatz:1994tz}
\bibinfo{author}{\bibfnamefont{S.~H.} \bibnamefont{Strogatz}},
  \emph{\bibinfo{title}{{Nonlinear dynamics and Chaos: with applications to
  physics, biology, chemistry, and engineering}}}
  (\bibinfo{publisher}{Westview}, \bibinfo{year}{1994}).

\bibitem[{\citenamefont{Masschaele et~al.}(2010)\citenamefont{Masschaele, Park,
  Furst, Fransaer, and Vermant}}]{Masschaele:2010da}
\bibinfo{author}{\bibfnamefont{K.}~\bibnamefont{Masschaele}},
  \bibinfo{author}{\bibfnamefont{B.~J.} \bibnamefont{Park}},
  \bibinfo{author}{\bibfnamefont{E.~M.} \bibnamefont{Furst}},
  \bibinfo{author}{\bibfnamefont{J.}~\bibnamefont{Fransaer}}, \bibnamefont{and}
  \bibinfo{author}{\bibfnamefont{J.}~\bibnamefont{Vermant}},
  \bibinfo{journal}{Phys. Rev. Lett.} \textbf{\bibinfo{volume}{105}},
  \bibinfo{pages}{048303} (\bibinfo{year}{2010}).

\bibitem[{sup()}]{suppmat}
\bibinfo{note}{See Supplemental Material at ??? for movies, details of methods
  and additional material characterization.}

\bibitem[{\citenamefont{Brooks et~al.}(1999)\citenamefont{Brooks, Fuller,
  Frank, and Robertson}}]{Brooks:1999ky}
\bibinfo{author}{\bibfnamefont{C.~F.} \bibnamefont{Brooks}},
  \bibinfo{author}{\bibfnamefont{G.~G.} \bibnamefont{Fuller}},
  \bibinfo{author}{\bibfnamefont{C.~W.} \bibnamefont{Frank}}, \bibnamefont{and}
  \bibinfo{author}{\bibfnamefont{C.~R.} \bibnamefont{Robertson}},
  \bibinfo{journal}{Langmuir} \textbf{\bibinfo{volume}{15}},
  \bibinfo{pages}{2450} (\bibinfo{year}{1999}).

\bibitem[{\citenamefont{Reynaert et~al.}(2008)\citenamefont{Reynaert, Brooks,
  Moldenaers, Vermant, and Fuller}}]{Reynaert:2008dm}
\bibinfo{author}{\bibfnamefont{S.}~\bibnamefont{Reynaert}},
  \bibinfo{author}{\bibfnamefont{C.~F.} \bibnamefont{Brooks}},
  \bibinfo{author}{\bibfnamefont{P.}~\bibnamefont{Moldenaers}},
  \bibinfo{author}{\bibfnamefont{J.}~\bibnamefont{Vermant}}, \bibnamefont{and}
  \bibinfo{author}{\bibfnamefont{G.~G.} \bibnamefont{Fuller}},
  \bibinfo{journal}{J. Rheol.} \textbf{\bibinfo{volume}{52}},
  \bibinfo{pages}{261} (\bibinfo{year}{2008}).

\bibitem[{\citenamefont{Kabla and Debr{\'e}geas}(2003)}]{Kabla:2003it}
\bibinfo{author}{\bibfnamefont{A.}~\bibnamefont{Kabla}} \bibnamefont{and}
  \bibinfo{author}{\bibfnamefont{G.}~\bibnamefont{Debr{\'e}geas}},
  \bibinfo{journal}{Phys. Rev. Lett.} \textbf{\bibinfo{volume}{90}},
  \bibinfo{pages}{258303} (\bibinfo{year}{2003}).

\bibitem[{\citenamefont{Picard et~al.}(2004)\citenamefont{Picard, Ajdari,
  Lequeux, and Bocquet}}]{Picard:2004cx}
\bibinfo{author}{\bibfnamefont{G.}~\bibnamefont{Picard}},
  \bibinfo{author}{\bibfnamefont{A.}~\bibnamefont{Ajdari}},
  \bibinfo{author}{\bibfnamefont{F.}~\bibnamefont{Lequeux}}, \bibnamefont{and}
  \bibinfo{author}{\bibfnamefont{L.}~\bibnamefont{Bocquet}},
  \bibinfo{journal}{Eur. Phys. J. E} \textbf{\bibinfo{volume}{15}},
  \bibinfo{pages}{371} (\bibinfo{year}{2004}).

\bibitem[{\citenamefont{Weaire and Rivier}(1984)}]{Weaire:1984gu}
\bibinfo{author}{\bibfnamefont{D.}~\bibnamefont{Weaire}} \bibnamefont{and}
  \bibinfo{author}{\bibfnamefont{N.}~\bibnamefont{Rivier}},
  \bibinfo{journal}{Contemp. Phys.} \textbf{\bibinfo{volume}{25}},
  \bibinfo{pages}{59} (\bibinfo{year}{1984}).

\bibitem[{\citenamefont{Zaccone and Scossa-Romano}(2011)}]{Zaccone:2011cm}
\bibinfo{author}{\bibfnamefont{A.}~\bibnamefont{Zaccone}} \bibnamefont{and}
  \bibinfo{author}{\bibfnamefont{E.}~\bibnamefont{Scossa-Romano}},
  \bibinfo{journal}{Phys. Rev. B} \textbf{\bibinfo{volume}{83}},
  \bibinfo{pages}{184205} (\bibinfo{year}{2011}).

\bibitem[{\citenamefont{Cort{\'e} et~al.}(2008)\citenamefont{Cort{\'e},
  Chaikin, Gollub, and Pine}}]{Corte:2008tp}
\bibinfo{author}{\bibfnamefont{L.}~\bibnamefont{Cort{\'e}}},
  \bibinfo{author}{\bibfnamefont{P.~M.} \bibnamefont{Chaikin}},
  \bibinfo{author}{\bibfnamefont{J.~P.} \bibnamefont{Gollub}},
  \bibnamefont{and} \bibinfo{author}{\bibfnamefont{D.~J.} \bibnamefont{Pine}},
  \bibinfo{journal}{Nat. Phys.} \textbf{\bibinfo{volume}{4}},
  \bibinfo{pages}{420} (\bibinfo{year}{2008}).

\bibitem[{\citenamefont{Richard et~al.}(2005)\citenamefont{Richard, Nicodemi,
  Delannay, Ribi{\`e}re, and Bideau}}]{Richard:2005dj}
\bibinfo{author}{\bibfnamefont{P.}~\bibnamefont{Richard}},
  \bibinfo{author}{\bibfnamefont{M.}~\bibnamefont{Nicodemi}},
  \bibinfo{author}{\bibfnamefont{R.}~\bibnamefont{Delannay}},
  \bibinfo{author}{\bibfnamefont{P.}~\bibnamefont{Ribi{\`e}re}},
  \bibnamefont{and} \bibinfo{author}{\bibfnamefont{D.}~\bibnamefont{Bideau}},
  \bibinfo{journal}{Nat. Mater.} \textbf{\bibinfo{volume}{4}},
  \bibinfo{pages}{121} (\bibinfo{year}{2005}).

\bibitem[{\citenamefont{Nguyen et~al.}(2011)\citenamefont{Nguyen, Darnige,
  Bruand, and Clement}}]{Nguyen:2011hm}
\bibinfo{author}{\bibfnamefont{V.~B.} \bibnamefont{Nguyen}},
  \bibinfo{author}{\bibfnamefont{T.}~\bibnamefont{Darnige}},
  \bibinfo{author}{\bibfnamefont{A.}~\bibnamefont{Bruand}}, \bibnamefont{and}
  \bibinfo{author}{\bibfnamefont{E.}~\bibnamefont{Clement}},
  \bibinfo{journal}{Phys. Rev. Lett.} \textbf{\bibinfo{volume}{107}},
  \bibinfo{pages}{138303} (\bibinfo{year}{2011}).

\bibitem[{\citenamefont{H{\'e}braud et~al.}(1997)\citenamefont{H{\'e}braud,
  Lequeux, Munch, and Pine}}]{Hebraud:1997ef}
\bibinfo{author}{\bibfnamefont{P.}~\bibnamefont{H{\'e}braud}},
  \bibinfo{author}{\bibfnamefont{F.}~\bibnamefont{Lequeux}},
  \bibinfo{author}{\bibfnamefont{J.-P.} \bibnamefont{Munch}}, \bibnamefont{and}
  \bibinfo{author}{\bibfnamefont{D.~J.} \bibnamefont{Pine}},
  \bibinfo{journal}{Phys. Rev. Lett.} \textbf{\bibinfo{volume}{78}},
  \bibinfo{pages}{4657} (\bibinfo{year}{1997}).

\bibitem[{\citenamefont{Chen et~al.}(2011)\citenamefont{Chen, Manning, Yunker,
  Ellenbroek, Zhang, Liu, and Yodh}}]{Chen:2011fa}
\bibinfo{author}{\bibfnamefont{K.}~\bibnamefont{Chen}},
  \bibinfo{author}{\bibfnamefont{M.~L.} \bibnamefont{Manning}},
  \bibinfo{author}{\bibfnamefont{P.~J.} \bibnamefont{Yunker}},
  \bibinfo{author}{\bibfnamefont{W.~G.} \bibnamefont{Ellenbroek}},
  \bibinfo{author}{\bibfnamefont{Z.}~\bibnamefont{Zhang}},
  \bibinfo{author}{\bibfnamefont{A.~J.} \bibnamefont{Liu}}, \bibnamefont{and}
  \bibinfo{author}{\bibfnamefont{A.~G.} \bibnamefont{Yodh}},
  \bibinfo{journal}{Phys. Rev. Lett.} \textbf{\bibinfo{volume}{107}},
  \bibinfo{pages}{108301} (\bibinfo{year}{2011}).

\end{thebibliography}

\end{document}